\documentclass[a4paper,fleqn,usenatbib]{mnras}
\usepackage{txfonts}
\usepackage[T1]{fontenc}
\usepackage{ae,aecompl}

\usepackage{graphicx}	

\newcommand{\degree}{\ensuremath{^\circ}}
\newcommand{\snr}{SNR~1987A }
\newcommand{\sn}{SN~1987A }
\newcommand{\snpos}{SN~1987A's }
\newcommand{\snnospace}{SN~1987A}
\newcommand{\snrnospace}{SNR~1987A}

\title[Spectral Modelling of SNR 1987A]{Low Radio Frequency Observations and Spectral Modelling of the Remnant of Supernova 1987A}
\author[J. R. Callingham et al.]{J.~R.~Callingham,$^{1,2,3}$\thanks{E-mail: j.callingham@physics.usyd.edu.au}
B.~M.~Gaensler,$^{4,1,3}$
G.~Zanardo,$^{5}$
L.~Staveley-Smith,$^{5,3}$ 
\newauthor
P.~J.~Hancock,$^{6,3}$
N.~Hurley-Walker,$^{6}$
M.~E.~Bell,$^{2,3}$ 
K.~S.~Dwarakanath,$^{7}$ 
\newauthor
T.~M.~O.~Franzen,$^{6}$
L.~Hindson,$^{8}$
M.~Johnston-Hollitt,$^{8}$
A.~Kapi\'nska,$^{5,3}$
B.-Q.~For,$^{5}$
\newauthor
E.~Lenc,$^{1,3}$
B.~McKinley,$^{9,3}$
J.~Morgan,$^{6}$
A.~R.~Offringa,$^{10,3}$
P.~Procopio,$^{9,3}$
\newauthor
R.~B.~Wayth,$^{6,3}$
C.~Wu$^{5}$
and Q.~Zheng$^{8}$ 
\\
$^{1}$Sydney Institute for Astronomy (SIfA), School of Physics, The University of Sydney, NSW 2006, Australia\\
$^{2}$CSIRO Astronomy and Space Science (CASS), Marsfield, NSW 2122, Australia\\
$^{3}$ARC Centre of Excellence for All-Sky Astrophysics (CAASTRO), Australia\\
$^{4}$Dunlap Institute for Astronomy \& Astrophysics, University of Toronto, Toronto, ON, M5S 3H4, Canada \\
$^{5}$International Centre for Radio Astronomy Research (ICRAR), The University of Western Australia, Crawley, WA 6009, Australia\\
$^{6}$International Centre for Radio Astronomy Research (ICRAR), Curtin University, Bentley, WA 6102, Australia\\
$^{7}$Raman Research Institute (RRI), Bangalore 560080, India \\
$^{8}$School of Chemical \& Physical Sciences, Victoria University of Wellington, Wellington 6140, New Zealand\\
$^{9}$School of Physics, The University of Melbourne, Parkville, VIC 3010, Australia\\
$^{10}$Netherlands Institute for Radio Astronomy (ASTRON), Dwingeloo, The Netherlands\\
}
\date{Accepted 17 June 2016. Received 17 June 2016; in original form 2 May 2016}

\pubyear{2016}

\begin{document}
\label{firstpage}
\pagerange{\pageref{firstpage}--\pageref{lastpage}}
\maketitle

\begin{abstract}

\noindent We present Murchison Widefield Array observations of the supernova remnant (SNR) 1987A between 72 and 230\,MHz, representing the lowest frequency observations of the source to date. This large lever arm in frequency space constrains the properties of the circumstellar medium created by the progenitor of \snr when it was in its red supergiant phase. As of late-2013, the radio spectrum of \snr between 72\,MHz and 8.64\,GHz does not show any deviation from a non-thermal power-law with a spectral index of $-0.74 \pm 0.02$. This spectral index is consistent with that derived at higher frequencies, beneath 100\,GHz, and with a shock in its adiabatic phase. A spectral turnover due to free-free absorption by the circumstellar medium has to occur below 72\,MHz, which places upper limits on the optical depth of $\leq$\,0.1 at a reference frequency of 72\,MHz, emission measure of $\lesssim$\,13,000\,cm$^{-6}$\,pc, and an electron density of $\lesssim$\,110\,cm$^{-3}$. This upper limit on the electron density is consistent with the detection of prompt radio emission and models of the X-ray emission from the supernova. The electron density upper limit implies that some hydrodynamic simulations derived a red supergiant mass loss rate that is too high, or a wind velocity that is too low. The mass loss rate of $\sim 5 \times 10^{-6}$\,$M_{\sun}$\,yr$^{-1}$ and wind velocity of 10\,km\,s$^{-1}$ obtained from optical observations are consistent with our upper limits, predicting a current turnover frequency due to free-free absorption between 5 and 60\,MHz. 

\end{abstract}


\section{Introduction}

Supernova 1987A (SN~1987A), discovered in the Large Magellanic Cloud (LMC) on 1987 February 23, was the brightest supernova seen from Earth since the invention of the telescope \citep{1987IAUC.4316....1K,1987IAUC.4338....1K,1987IAUC.4340....1S}. The close proximity of \snnospace, and the detailed information we have about its progenitor, has meant \sn has played a pivotal role in shaping our understanding of core-collapse supernovae, supernova remnant (SNR) evolution, and the physical properties of the circumstellar medium deposited by a supernova progenitor.

Radio emission from core-collapse supernovae generally occurs when the forward shock sweeps up the dense, slow-moving wind generated by a red supergiant progenitor \citep{1982ApJ...259..302C}. However, the progenitor to \sn was very different than those normally associated with radio supernovae. It is believed that \snpos progenitor, Sk-69\degree202, evolved from a red supergiant into a blue supergiant $\sim$\,20,000 years before the supernova event \citep{1991Natur.350..683C,2007AIPC..937..125P}. While the cause of such a transformation is still debated in the literature \citep[e.g.][]{1987ApJ...318..664W,1989Natur.338..401P,1999ApJ...512..322C,2007AJ....133.1034S}, this abnormal evolutionary path for a core-collapsed supernova progenitor has produced a complex environment which the shock from the supernova event interacts with. 

For example, the prompt radio emission detected at 843\,MHz by the Molonglo Observatory Synthesis Telescope (MOST) \citep{1987Natur.327...38T}, which faded to an undetectable flux density in under a year \citep{2001ApJ...549..599B}, is understood to be a consequence of the supernova shock interacting with the low density, fast moving blue supergiant wind \citep{1987Natur.329..421S}. Radio emission was then re-detected $\sim$1200 days after the core collapse \citep{1993Natur.366..136S}, indicating that the forward shock of the supernova event was encountering the denser and slower red supergiant wind in the equatorial plane. This interaction produced the radio shell that is now colliding with the clumpy ring observed at optical wavelengths \citep{1995ApJ...438..724C,1995ApJ...439..730P}. The hourglass shape of \snrnospace, formed from the peculiar evolution of the progenitor \citep{1995ApJ...452L..45C,2005ApJ...627..888S,2014ApJ...794..174P}, also means the forward shock is interacting with hot blue supergiant wind material beyond the termination region at high latitudes \citep{2010ApJ...710.1515Z}. Currently, $>$\,9,600 days since the supernova event, it is understood that the forward shock has egressed the equatorial ring and is now interacting with the H\,\textsc{ii} and hourglass region; a region formed from the blue supergiant wind expanding into the red supergiant wind \citep{1999ApJ...511..389L,2014ApJ...794..174P}. 


While \snr has been extensively studied across the electromagnetic spectrum, there have been no observations conducted at low radio frequencies ($< 0.843$\,GHz). This was due to \snr being too low in declination to be observed with Northern Hemisphere radio telescopes, and because \sn occurred after all the low radio frequency instruments in the Southern Hemisphere were decommissioned, such as the Culgoora circular array \citep{1995AuJPh..48..143S}. Low frequency radio observations of SNRs can constrain the radio spectrum, providing a unique probe for investigating the circumstellar medium via free-free absorption and insights into the shock acceleration process producing the radio emission \citep[e.g.][]{1989ApJ...347..915K,1995ApJ...455L..59K,2001ApJ...559..954L,2005AJ....130..148B}. Since intrinsic or extrinsic spectral variations are often subtle, low radio frequency observations are an important key in obtaining a large enough lever arm in frequency space to identify any variation. In particular, low radio frequency observations can place constraints on the electron density of the absorbing medium and the mass loss rate of the progenitor \citep{1982ApJ...259..302C,1989ApJ...347..915K,1990LNP...362..130C,2014ApJ...785....7D}. Considering the forward shock of \snr has now passed through the densest part of the equatorial ring \citep{2014ApJ...794..174P,2015ApJ...806L..19F}, low frequency observations of \snr can provide mass loss limits of the progenitor when it was in its red supergiant phase. This is because the detected radio emission from \snr is dominated by emission from the region between the forward shock, that is propagating into the circumstellar material, and the reverse shock \citep{1982ApJ...259..302C}. The temperature and density of the material internal to the reverse shock is such that the radio emission from the interior to the reverse shock is completely absorbed \citep{1982ApJ...259..302C,1999ApJ...511..389L,2003LNP...598..171C}.

In this paper we present the lowest radio frequency observations of \snr using the Murchison Widefield Array \citep[MWA;][]{Tingay2013}. The MWA is a low radio frequency aperture array which observed \snr between 72 and 231\,MHz as part of its all-sky survey \citep{2015PASA...32...25W}. The observations presented in this paper are over an order of a magnitude lower in frequency than the previous lowest-frequency observations of \snrnospace, allowing us to investigate the surrounding circumstellar medium in a part of frequency space previously unobservable. Combined with gigahertz observations from the Australia Telescope Compact Array (ATCA), these observations provide key insights into the interaction of the supernova shock with the circumstellar medium and its properties. The MWA and the ATCA observations, and the relevant data reduction procedures performed, are outlined in detail in \S\,\ref{sec:obs}. The resulting radio spectrum and the absorption model fits are described in \S\,\ref{sec:results}. In \S\,\ref{sec:discuss}, the implications of these model fits on the mass-loss rate and wind velocity of the progenitor of \snr are discussed.

\section{Observations and Data Reduction}
\label{sec:obs}

\snr was observed by the MWA and the ATCA in late 2013 and early 2014. While not simultaneous, the two ATCA observations bracket the MWA observation with an almost equal gap of three months either side. These observations are summarised in Table \ref{tab:flux_den} and the details about the data reduction processes are described below.

\begin{table}
	\small
	\caption{\label{fluxvals} A summary of the observations of \snr used in the spectral modelling. Note that $\nu$ represents the central frequency of the observing band, $S_{\nu}$ is the total flux density at frequency $\nu$, and $\Delta S_{\nu}$ is the uncertainty on the flux density measurement. For the MWA measurements $\Delta S_{\nu}$ is the sum in quadrature of the local RMS noise and the systematic uncertainties associated with correcting the deficiencies in the primary beam model. $\Delta S_{\nu}$ for the ATCA measurements is the sum in quadrature of the local RMS noise and the uncertainties in the gain calibration. $a_{\mathrm{PSF}}$ and $b_{\mathrm{PSF}}$ are the semi-major and semi-minor axis of the synthesised beam, respectively.}
	\label{tab:flux_den}
	\begin{center}
		\begin{tabular}{ccccccc}
		\hline
		\hline
$\nu$ & $S_{\nu}$ & $\Delta S$ & Epoch & Telescope & $a_{\mathrm{PSF}}$ & $b_{\mathrm{PSF}}$\\
(GHz) & (Jy) & (Jy) & & & (\arcmin) & (\arcmin) \\
		\hline			
0.076 & 5.1       & 0.8  & 2013 Nov 08 & MWA & 5.9 & 5.2 \\
0.084 & 4.9       & 0.7  & 2013 Nov 08 & MWA & 5.4 & 4.8 \\
0.092 & 4.7       & 0.5  & 2013 Nov 08 & MWA & 5.0 & 4.4 \\
0.099 & 4.6       & 0.4  & 2013 Nov 08 & MWA & 4.8 & 4.1 \\
0.107 & 4.5       & 0.3  & 2013 Nov 08 & MWA & 4.3 & 3.7 \\
0.115 & 4.2       & 0.2  & 2013 Nov 08 & MWA & 4.0 & 3.5 \\
0.123 & 4.0       & 0.2  & 2013 Nov 08 & MWA & 3.8 & 3.3 \\
0.130 & 3.9       & 0.2  & 2013 Nov 08 & MWA & 3.7 & 3.1 \\
0.143 & 3.6       & 0.2  & 2013 Nov 08 & MWA & 3.4 & 2.8 \\
0.150 & 3.4       & 0.1  & 2013 Nov 08 & MWA & 3.2 & 2.6 \\
0.158 & 3.3       & 0.1  & 2013 Nov 08 & MWA & 3.1 & 2.5 \\
0.166 & 3.1       & 0.1  & 2013 Nov 08 & MWA & 3.0 & 2.4 \\
0.174 & 3.0       & 0.1  & 2013 Nov 08 & MWA & 2.8 & 2.3 \\
0.181 & 2.9       & 0.1  & 2013 Nov 08 & MWA & 2.7 & 2.2 \\
0.189 & 2.8       & 0.1  & 2013 Nov 08 & MWA & 2.6 & 2.1 \\
0.197 & 2.7       & 0.1  & 2013 Nov 08 & MWA & 2.5 & 2.1 \\
0.204 & 2.5       & 0.1  & 2013 Nov 08 & MWA & 2.4 & 2.0 \\
0.212 & 2.5       & 0.1  & 2013 Nov 08 & MWA & 2.3 & 1.9 \\
0.219 & 2.4       & 0.1  & 2013 Nov 08 & MWA & 2.3 & 1.8 \\
0.227 & 2.3       & 0.1  & 2013 Nov 08 & MWA & 2.2 & 1.7 \\
1.375 & 0.58      & 0.05 & 2013 Aug 31 & ATCA & 0.11 & 0.09\\
1.375 & 0.58      & 0.04 & 2014 Feb 04 & ATCA & 0.08 & 0.07\\
2.351 & 0.43      & 0.02 & 2013 Aug 31 & ATCA & 0.06 & 0.05\\
2.351 & 0.42      & 0.02 & 2014 Feb 04 & ATCA & 0.05 & 0.04\\
4.788 & 0.28      & 0.01 & 2013 Aug 31 & ATCA & 0.03 & 0.02\\
4.788 & 0.30      & 0.02 & 2014 Feb 04 & ATCA & 0.03 & 0.02\\
8.642 & 0.18      & 0.03 & 2013 Aug 31 & ATCA & 0.01 & 0.01\\
8.642 & 0.17      & 0.02 & 2014 Feb 04 & ATCA & 0.01 & 0.01 \\                                                 
\hline\end{tabular}                           
\end{center}                                                                               
\end{table}

\subsection{MWA Observations and Data Reduction}

\snr was observed by the MWA between 72 and 231\,MHz on 2013 November 8 as part of the GaLactic and Extragalactic All-sky Murchison Widefield Array \citep[GLEAM;][]{2015PASA...32...25W} survey. The GLEAM survey was conducted by observing the sky between declinations of $-$90\degree and $+$25\degree in a two-minute ``snapshot'' mode, utilising the meridian drift scan technique at seven independent declination settings. \snr was observed during the drift scan that was centred on a declination of $-55$\degree. 

The data reduction process that was performed is described in detail by \citet{2015PASA...32...25W} and Hurley-Walker et al. (submitted). In summary, \textsc{Cotter} \citep{2015PASA...32....8O} was used to process the raw visibility data from the MWA observations, which involved averaging the data to 1\,s time and 40\,kHz frequency resolution, and excising radio frequency interference (RFI) using the \textsc{AOFlagger} algorithm \citep{2012A&A...539A..95O}. An initial model of the sky for the five instantaneous observing bandwidths of 30.72\,MHz was produced by observing bright calibrator sources. Hydra A was the calibrator source observed for the declination strip of the survey that included \snrnospace. Such observations allowed initial amplitude and phase calibration solutions to be applied. 

\textsc{WSClean} \citep{2014MNRAS.444..606O} was used for imaging as it appropriately accounts for wide-field $w$-term effects. \textsc{WSClean} is a fast generic widefield imager that uses the $w$-stacking algorithm \citep{Humphreys2013} and the $w$-snapshot algorithm \citep{2012SPIE.8500E..0LC}. Since \snr is unresolved at all MWA observing frequencies, baselines shorter than 60\,m were excluded to minimise the contamination of large scale, diffuse structure present in the LMC, and a $u,v$-weighting scheme close to uniform weighting was chosen for imaging. In terms of the ``Briggs'' scheme this corresponds to a ``robust'' parameter of $-1.0$ \citep{Briggs1995}. The snapshot observations were imaged across each 30.72\,MHz band using multi-frequency synthesis down to the first negative \textsc{clean} component, without any major cycles. These images then went through a self-calibration loop, using the initial calibration images for quality control and to ensure the position of sources and the flux density of the image were consistent throughout the process. The RMS of the image was then measured and a new \textsc{clean} threshold was set to three times that RMS, which was between 200 to 40\,mJy for 72 to 231\,MHz, respectively. The observations were then divided into four 7.68-MHz sub-bands and jointly cleaned. 

The Molonglo Reference Catalogue \citep[MRC;][]{Large1981,1991Obs...111...72L} was used to set an initial flux density scale for the image and to correct for any right ascension dependent flux density scale errors due to the drift scan technique. An astrometric correction was also performed at this stage to fix bulk ionospheric distortions by using the positions of sources referenced in MRC. The snapshots for the entire observed declination strip were then mosaicked, with each snapshot weighted by the square of the primary beam response. 

Finally, the residual declination dependence in the flux density scale, due to uncertainties in the primary beam model, was corrected. This was done by comparing the measured flux density in the mosaics for isolated, unresolved sources above 8$\sigma$ of the noise floor to that predicted by their radio spectra using the 74\,MHz Very Large Array Low-Frequency Sky Survey Redux \citep[VLSSr;][]{2014MNRAS.440..327L}, 408\,MHz MRC, and the 1.4\,GHz NRAO VLA Sky Survey \citep[NVSS;][]{Condon1998}. This correction method places the flux density measurements on the \citet{Baars1977} flux density scale and dominates the uncertainty in the flux density measurements. The flux density calibration is accurate to 8\%, as assessed by comparing the MWA flux densities at 150\,MHz and 230\,MHz to the 150\,MHz TGSS-ADR1 survey \citep{2016arXiv160304368I} and Jansky Very Large Array (JVLA) P-band (230-450\,MHz) observations of compact, non-variable sources (Hurley-Walker et al., submitted).

\begin{figure}
    \includegraphics[scale=0.36]{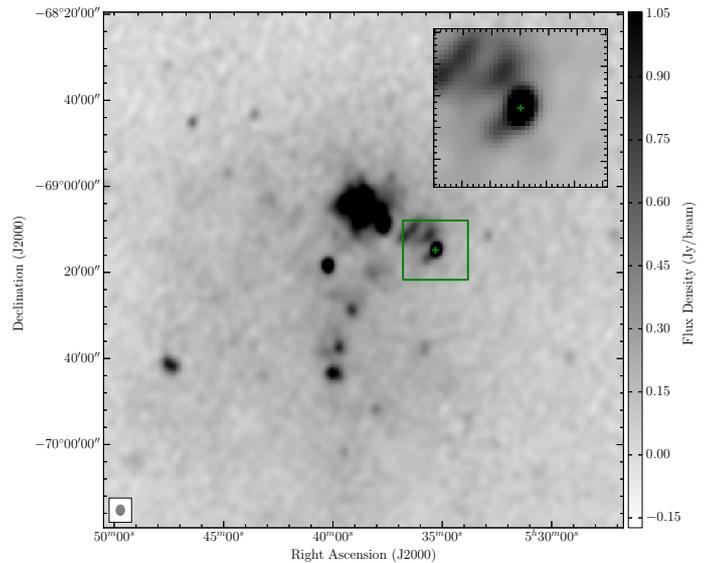}
    \caption{A MWA image of a part of the LMC at 200-208 MHz formed using a robust weighting parameter of $-1$ (close to uniform weighting). The position of SNR~1987A is marked by a dark green cross. The inset is a magnified section of the larger image, as represented by the dark green box. The dimensions of the inset are 5$\farcm$6 $\times$ 5$\farcm$6. In the top left of the inset is the superbubble 30 Doradus C \citep{1984AuJPh..37..321M,2015A&A...583A.121K} and the source to the south-east of SNR~1987A is a blend of the SNR called the Honeycomb nebula \citep{1995AJ....109.1729C} and a background galaxy \citep{2001ApJ...549..599B}. The synthesised beam size for this observation is 2$\farcm$4 $\times$ 2$\farcm$0, and is plotted in the bottom left hand corner.}
    \label{fig:cutout}
\end{figure}

We convolved the appropriate synthesised beam at each sub-band frequency to characterise the flux density of all the sources within two degrees of the centre of the LMC for each of the 20 sub-band images. The sub-band image at 200-208 MHz, with \snr highlighted, is shown in Figure \ref{fig:cutout}. 

The background emission and noise properties of the individual sub-band images were measured using the backgrounding tool \textsc{Background And Noise Estimator (BANE)}\footnote{\texttt{https://github.com/PaulHancock/Aegean/wiki/BANE}}. \textsc{BANE} defines the background to be the mean of the pixel distribution, and the noise to be the variance about this mean. \textsc{BANE} was designed to quickly and accurately treat some of the unique problems of estimating the background and noise properties of radio images. It utilises two main techniques to reduce the compute time, whilst retaining a high level of accuracy. Firstly, since radio images can have a high level of correlation between adjacent pixels, \textsc{BANE} calculates the mean and variance on a sparse grid of pixels and then interpolates to give the final background and noise images. Secondly, \textsc{BANE} uses sigma-clipping on the pixel distribution to avoid contamination from source pixels.

\textsc{BANE} calculated the background emission of the MWA images to vary between $\sim$0.7 Jy/beam to $\sim$0.1 Jy/beam for 72 to 231\,MHz, respectively. The background and noise properties were then used by the source finding and characterisation program \textsc{Aegean} v1.9.6 \citep{2012MNRAS.422.1812H} to accurately identify and measure the flux density of the sources in the images. \snr was unresolved in all the subband images, so the flux density measurements were calculated by \textsc{Aegean} by fitting a Gaussian, convolved with the synthesised beam, to the source position.

At the lowest four frequencies, the synthesised beam becomes large enough to cause some blending of the Honeycomb nebula \citep{1995AJ....109.1729C} and a background galaxy \citep{2001ApJ...549..599B} with \snrnospace, shown to the south-east of \snr in the inset of Figure \ref{fig:cutout}. We estimate the upper limit of the contamination at these frequencies by extrapolating the spectra of the Honeycomb nebula and the background galaxy from a fit above the lowest four frequencies of the MWA data. This results in larger uncertainties for the flux density measurements between 72 and 103 MHz of \snrnospace.

\subsection{ATCA Observations and Data Reduction}

\snr was observed as part of the ongoing ATCA monitoring campaign (project C015, PI Staveley-Smith) on 2013 August 31 and 2014 February 4 in array configurations 1.5A and 6D, respectively. The observations were conducted using the Compact Array Broadband Backend \citep[CABB;][]{Wilson2011} system, providing an instantaneous 2\,GHz bandwidth, for both linear polarisations, at central frequencies of 2.1\,GHz, 5.5\,GHz and 9.0\,GHz. For both observations PKS~B1934-638 was used for gain and bandpass calibration and to set the flux density scale. PKS~B0454-810, PKS~B0407-658 and PKS~B0530-727 were the secondary calibrators used for phase calibration. The total integration time on \snr in the August and February observations was approximately seven and six hours, respectively. 

The data reduction process we applied is outlined by \citet{2010ApJ...710.1515Z}. In summary, the data for both observations were reduced using the \textsc{miriad} software package \citep{Sault1995}, with known regions of RFI and lower sensitivity in the CABB system initially flagged. The excision of RFI was conducted using the automatic flagging option in $\tt{pgflag}$ and manually with $\tt{blflag}$. Since the LMC is a crowded field with many bright sources nearby \snrnospace, only baselines longer than 3\,k$\lambda$, where $\lambda$ is the observing wavelength, were used to form images. To be consistent with pre-CABB monitoring data, the 2\,GHz instantaneous bandwidth was spilt into four sub-bands with 128\,MHz bandwidth, centred at 1.4, 2.4, 4.8 and 8.6\,GHz. Gain calibration was performed on each sub-band independently. A self-calibration loop was conducted at this stage using a preliminary model generated by \textsc{clean}ing an image with a small number of iterations, with deeper \textsc{clean}ing conducted after the self-calibration loop was complete. The flux densities were measured by integrating a Gaussian fit to the emission region at 1.4 and 2.4\,GHz, and by integrating over a polygonal region at 4.8 and 8.6\,GHz.

\section{Results}
\label{sec:results}

\begin{figure*}
	\includegraphics[scale=0.37]{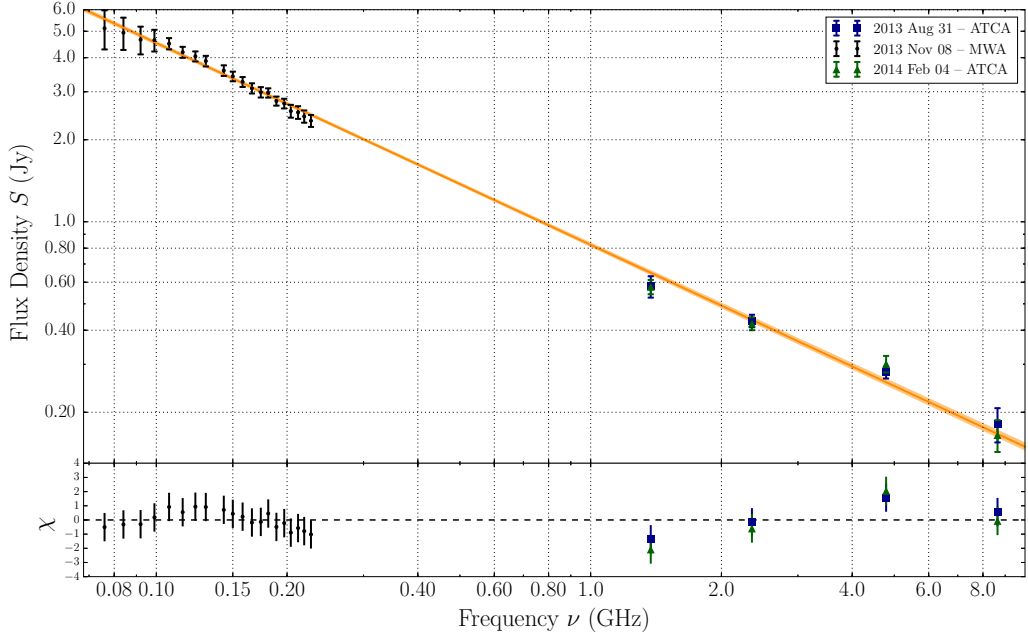}
    \caption{Spectral energy distribution of \snr from 0.072 to 8.64\,GHz. The MWA data points are plotted in black. The monitoring data from the ATCA for 2013 August 31 and 2014 February 4 are shown in blue and green, respectively. The best fit power-law model to all the data is presented in orange, with the shaded orange region representing the 1-$\sigma$ uncertainty on the model fit at the respective frequency. The $\chi$-values for the power-law fit to the data, which represent the residuals to the fit divided by the uncertainties in the flux density measurements, are displayed in the panel below the spectral energy distribution.}
    \label{fig:sed}
\end{figure*}

We present the spectral energy distribution of \snr from 72\,MHz to 8.64\,GHz in Figure \ref{fig:sed}, plotted using the values reported in Table \ref{tab:flux_den}. Applying the Bayesian model inference routine outlined in \citet{Callingham2015}, different emission and absorption models were fit to find the model that best described the spectrum and to test for any evidence of a spectral turnover or broad spectral curvature. This fitting method assumes that the flux density measurements are Gaussian and the ATCA data points are independent. The known correlation in the MWA flux measurements (Hurley-Walker et al., submitted) was approximated by a Mat\'{e}rn covariance function \citep{Rasmussen2006}, which produces a stronger correlation between flux density measurements close in frequency space than further away. The Bayesian evidence $Z$ for each fit was calculated using the using the algorithm \textsc{MultiNest} \citep{Feroz2013}, which is an implementation of nested sampling. Uniform priors for each model parameter in a fit were utilised, allowing direct comparison to least-squares goodness-of-fit tests. While the spectral index of \snr has been gradually flattening with time \citep{2010ApJ...710.1515Z}, the data were simultaneously fit since the ATCA observations were almost equally spaced before and after the MWA observation. While the flattening of the spectral index is small over the time between the ATCA and MWA observations ($\approx$\,0.005), fitting the data simultaneously minimises any impact of variability.

The physically motivated models investigated included non-thermal synchrotron emission from relativistic electrons at the forward shock front, and homogeneous free-free absorption \citep{Mezger1967} caused by the ionised circumstellar material swept up by the shock front. The homogeneous free-free absorption model includes attenuation of the underlying non-thermal synchrotron radiation by an ionised screen internal or external to the emitting electrons. We found the best fitting model to be the non-thermal synchrotron emission of the form:

\begin{equation}\label{eqn:powlaw}
 S_{\nu} = a\left(\frac{\nu}{1\,\mathrm{GHz}}\right)^{\alpha},
\end{equation} 

\noindent where $a$, in Jy, characterises the amplitude of the synchrotron spectrum, $\alpha$ is the synchrotron spectral index, and $S_{\nu}$ is the flux density at frequency $\nu$, in GHz. The best fit of this model requires $\alpha = -0.74 \pm 0.02$ and $a = 0.82 \pm 0.01$ Jy. The fit is plotted in Figure \ref{fig:sed} and has a log evidence $\ln(Z)$ value of $8.32 \pm 0.02$, or a reduced $\chi^{2}$-value of 0.84, calculated using 25 degrees of freedom. This spectral index is somewhat steeper, but still consistent, with what is expected from the higher frequency ATCA data and leads to a shock compression ratio of $\sigma_{\mathrm{s}} = 3.02 \pm 0.04$, consistent with the range of compression ratios derived for \snr between 1.4 and 44\,GHz \citep{2006ApJ...650L..59B,2010ApJ...710.1515Z,2014ApJ...796...82Z}. This low compression factor implies the shock is still in the adiabatic phase and that sub-diffusive shock acceleration, without cosmic-ray feedback, is present.

While the spectrum of \snr is best described by a non-thermal power-law, we also fit an extrinsic free-free absorption model to place an upper limit on the optical depth. Assuming the ionised material is not mixed with the relativistic electrons that are producing the non-thermal spectrum, a spectrum with a peak below 72\,MHz is characterised as

\begin{equation}\label{eqn:homobremss}
	S_{\nu} = a \left(\frac{\nu}{0.072\,\mathrm{GHz}}\right)^{\alpha} \exp\left[-\tau_{72}\left(\frac{\nu}{0.072\,\mathrm{GHz}}\right)^{-2.1}\right],
\end{equation}

\noindent where $\tau_{72}$ is the free-free optical depth at the reference frequency of 72\,MHz. The fit to the spectrum requires $\tau_{72} \leq 0.1$ at $3\sigma$. 

Model selection can be performed based on the difference between the log evidence of two models $\Delta\ln(Z) = \ln(Z_{2}) - \ln(Z_{1})$, where $\Delta\ln(Z) \geq 3$ is taken as strong evidence that the second model is favoured over the first \citep{Kass1995}. The difference in the log evidence value between the free-free absorption fits and synchrotron radiation fit were found to be greater than 200, implying that the non-thermal synchrotron emission is strongly favoured over both internal and external free-free absorption. Non-physically motivated models, such as quadratic and quartic curves, were also fit to the spectrum to test for spectral curvature. The difference in evidence between these non-physical models and the synchrotron spectrum was always greater than 165, suggesting there is no statistical evidence of curvature in the spectrum of \snrnospace. This implies that if a turnover exists in the spectrum of \snrnospace, it has to occur at a frequency lower than 72\,MHz. 

Note that we excluded synchrotron self-absorption as a potential absorption mechanism as it was shown by \citet{1998ApJ...499..810C} that synchrotron self-absorption ceased almost immediately after the prompt burst. Additionally, the Razin-Tsytovich effect \citep{Tsytovich1951,Razin1957} is unlikely to be contributing to the absorption because it would require the density of the radiating region to be nearly two orders of magnitude larger than the postshock density, which is greater than expected for the synchrotron emitting region of \snr \citep{1982ApJ...259..302C,1995ApJ...452L..45C}. Therefore, free-free absorption by an ionised circumstellar material is the only plausible mechanism for a spectral turnover, considering the observations reported in this paper were conducted over 9640 days since \snr occurred.

\section{Discussion}
\label{sec:discuss}

The MWA observation of \snr represents the lowest frequency detection of \snrnospace, over an order of magnitude lower than the previous lowest frequency observations at 843\,MHz. This allows us to place a limit on the density of ionised material present around the shock producing the synchrotron emission. Since the best fitting model is a non-thermal power-law, if a turnover is present in the spectrum, it has to occur at or below a frequency of 72\,MHz. The medium responsible for the ionised material would likely be the surrounding H\,\textsc{ii} and ``hourglass'' regions deposited by the expansion of \snr progenitor's wind during its red supergiant phase \citep{1995ApJ...452L..45C}. This is because the forward shock front has swept around the dense equatorial ring $\sim$1000 days before the observations presented in this paper were made \citep{2014ApJ...794..174P}. 

Note that our observations are dominated by the emission from the expanding synchrotron emitting shell produced from the interaction of the forward shock with the circumstellar environment \citep{1982ApJ...259..302C}. The radio emission interior to this shell is not detected since the density in this region is too low for efficient particle acceleration environments to form, and any emission is absorbed by the dense material that exists between the reverse shock and the interior of the supernova \citep{1999ApJ...511..389L,2003LNP...598..171C,2014ApJ...794..174P}. Therefore, our observations are only sensitive to absorption by the circumstellar material deposited by the progenitor, rather than intrinsic absorption by the supernova ejecta.

To place an upper limit on the emission measure $\mathrm{EM} = \int fn^{2}_{\mathrm{e}}~\mathrm{d}l$, we re-parametrise the free-free optical depth from Equation \ref{eqn:homobremss} in terms of $T_{\mathrm{e}}$ in K, free electron density $n_{\mathrm{e}}$ in cm$^{-3}$, the path length $l$ in pc, and filling factor of the ionised gas $f$ as

\begin{equation}\label{eqn:opticaldepth}
	\tau_{\nu} \approx 8.24 \times 10^{-2} a(T_{\mathrm{e}},\nu) \nu^{-2.1} T_{\mathrm{e}}^{-1.35} \int fn^{2}_{\mathrm{e}}~\mathrm{d}l.
\end{equation} 

\noindent The Gaunt factor $a(T_{\mathrm{e}},\nu) \approx 1$ for the range of astrophysical quantities investigated in this study \citep{1989agna.book.....O}. The form of Equation \ref{eqn:opticaldepth} is derived assuming the H\,\textsc{ii} region is composed of hydrogenic gas, implying an atomic number of $\sim\,$1, and the number density of electrons and ions will be equal since the H\,\textsc{ii} region will be fully ionised from the ultraviolet (UV) radiation emitted from the supernova event. Emission line measurements and environmental simulations suggest the electron temperature of the H\,\textsc{ii} and ``hourglass'' regions will be $\sim 6 \times 10^{4}$\,K after the UV flash \citep{1991ApJ...380..575L,1999ApJ...511..389L,2005ApJ...627..888S} and before the forward shock of the supernova passes through \citep{1995ApJ...452L..45C,2014ApJ...794..174P}. Substituting this information, and the fact $\tau_{72} \leq 0.1$, into Equation \ref{eqn:opticaldepth} places an upper limit on the emission measure of $\mathrm{EM} \lesssim 13,000$\,cm$^{-6}$\,pc. 

While there is ambiguity about whether the turnover in the spectrum of Galactic SNRs is intrinsic or due to the ionised interstellar medium of the Milky Way \citep{1989ApJ...347..915K}, the path length of a photon from \snr through the interstellar medium of the LMC is significantly shorter than most Galactic SNRs. For example, since the electron density of the LMC near the position of \snr is measured to be $n_{\mathrm{e}} \sim 0.08$\,cm$^{-3}$ \citep{2003ApJS..148..473K,2006A&A...447..991C} and the distance of \snr from the edge of the LMC is $\sim 1$\,kpc \citep{1995ApJ...451..806X,2005ApJ...627..888S}, the interstellar medium of the LMC has an emission measure $\mathrm{EM} \lesssim 6$\,cm$^{-6}$\,pc. Hence, any turnover in the spectrum of \snr would be completely dominated by absorption from material associated with the system of \snrnospace. 

The emission measure also allows us to place an upper limit on the electron density in the red supergiant wind. For a radio photon emitted from the forward shock it will have an equivalent path length $l \sim$\,1\,pc, based on the current position of the forward shock and the total size of the H\,\textsc{ii} and ``hourglass'' region, assuming the progenitor was in the red supergiant phase for $\sim 5 \times 10^{5}$\,yr \citep{1993ApJ...405..337B,1995ApJ...452L..45C,2014ApJ...794..174P}. Provided the ionised material is distributed in a slab with a uniform density ($f = 1$), this produces an upper limit on the electron density $n_{\mathrm{e}} = \sqrt{\mathrm{EM}/fl} \lesssim 110$\,cm$^{-3}$. While the filling factor of the H\,\textsc{ii} and ``hourglass'' region is unknown, any deviation from unity will be small \citep[e.g.][]{1999ApJ...511..389L,2008A&A...484..371O,2015ApJ...806L..19F}, and will have a minimal impact on the electron density limit since dependency on the filling factor goes as $n_{\mathrm{e}} \propto f^{-0.5}$. 

An electron density $n_{\mathrm{e}}\lesssim 110$\,cm$^{-3}$ is consistent with the limits placed from the detection of the prompt radio emission, as any circumstellar envelope must be relatively thin otherwise the prompt radio burst would not have been observed \citep{1987Natur.329..421S}. Such an upper limit on the electron number density is also compatible with models of the X-ray emission from the supernova, which often require a electron density of $n_{\mathrm{e}} \sim 90$\,cm$^{-3}$ \citep{1997ApJ...477..281B,2002ApJ...567..314P} in the H\,\textsc{ii} and ``hourglass'' region.

We can also estimate the density of the H\,\textsc{ii} and ``hourglass'' region from the mass loss the progenitor underwent when it was a red supergiant. The asymmetric wind profile from \citet{1993ApJ...405..337B} has the environmental mass density $\rho$ described in terms of the mass loss rate of the red supergiant $\dot{M}$, velocity of the red supergiant wind $v_{\mathrm{w}}$, radius from centre of the system $r$, and the asymmetry parameter $A$ as

 \begin{equation}\label{eqn:massdensity}
 \rho(r,\theta) = \frac{3\dot{M}}{(3-A) 4 \pi r^{2} v_{\mathrm{w}}}(1 - A\cos^{2}\theta),
\end{equation} 

\noindent where $\theta$ is the angle from the pole of the progenitor. The best fitting model from the two dimensional hydrodynamic simulations of \citet{1993ApJ...405..337B} found $\dot{M} = 2.0 \times 10^{-5}$\,$M_{\sun}$\,yr$^{-1}$, $v_{\mathrm{w}} = 5$\,km\,s$^{-1}$ and $A = 0.95$. The asymmetry of the mass loss produces an equatorial to polar density ratio of 20:1. The largest number density of electrons, and thus the most likely site of absorption, will occur just before the forward shock front, which at day 9640 is at $r \sim 0.4$\,pc \citep{2014ApJ...794..174P}. In the equatorial plane this model predicts $n_{\mathrm{e}} \approx 120$\,cm$^{-3}$. This value is close to, but above, the limit we place based on the MWA observations of \snrnospace.

Therefore, the number density of electrons derived from the model of \citet{1993ApJ...405..337B}, which is also used by \citet{2014ApJ...794..174P} to model the initial environment of the system, is marginally inconsistent with the upper limit placed by our observations. We note that the inconsistency between the derived and observed electron density is highly dependent on the value of $r$. To ensure our findings are robust, we chose a value of $r$ that was at the upper-limit of those calculated from the simulations of \citet{2014ApJ...794..174P} and the observations of \citet{2014ApJ...796...82Z}. Since we have used a conservatively large value of $r$, it is likely either the mass loss rate is too high and/or the red supergiant wind velocity is too low in the model of \citet{1993ApJ...405..337B}.

Since the mass loss rate and wind speed are degenerate, we can parametrise the optical depth in terms of the physical properties of the supernova shock to place a limit on the ratio of the progenitor's mass loss rate and wind speed. Using the velocity of the forward shock $v_{\mathrm{s},4}$, in units of 10$^{4}$\,km\,s$^{-1}$, and the time since the explosion $t_{7}$, in units of 10$^{7}$\,s, the ratio of the mass loss rate and velocity of the wind can be expressed as

 \begin{equation}\label{eqn:masslossandwind}
 \left(\frac{\dot{M}_{-5}}{v_{\mathrm{w},1}}\right)^{2} = 0.25 \nu^{2} \tau_{\nu}  (v_{\mathrm{s},4} t_{7})^{3},
\end{equation} 

\noindent where the mass loss rate is in units of $10^{-5}$\,$M_{\sun}$\,yr$^{-1}$, the velocity of the red supergiant wind is in units of 10\,km\,s$^{-1}$, and the frequency is in units of GHz \citep{1981ApJ...251..259C,1982ApJ...259..302C}. Since the observations reported in this paper were conducted $\approx$\,9640 days since SN~1987A occurred, with $\tau_{72} \leq 0.1$, and $v_{\mathrm{s}} \sim 5000$\,km\,s$^{-1}$ \citep{2014ApJ...794..174P}, it follows that $\dot{M}_{-5}/v_{\mathrm{w},1} \leq 2.2$\,$M_{\sun}$\,yr$^{-1}$\,km$^{-1}$\,s. Again, the mass loss rate and wind velocity derived by \citet{1993ApJ...405..337B} are not consistent with our observations. Note that Equation \ref{eqn:masslossandwind} may lead to an overestimate of $\dot{M}_{-5}/v_{\mathrm{w},1}$ if the medium is found to be significantly clumpy \citep{2003LNP...598..171C}.

Using optical and infrared observations of \snrnospace, \citet{2005ApJ...627..888S} derived a mass loss rate of $\dot{M} \sim 5 \times 10^{-6}$\,$M_{\sun}$\,yr$^{-1}$, which is close to the median mass loss rate of the red supergiant population \citep{1985ApJ...292..640K}. Additionally, \citet{1992ApJ...388...93K} suggest a more realistic wind velocity of $v_{\mathrm{w}} = 10$\,km\,s$^{-1}$, corresponding to a ``slow wind'' expanding into a tenuous medium and becoming immediately radiative. Applying either, or both, of these values to Equation \ref{eqn:masslossandwind} provides a mass loss rate-wind velocity ratio consistent with our upper limit, and predicts a spectral turnover frequency between $\sim$\,5 and 60\,MHz. 

Future observations of \snr at or below 50\,MHz could be helpful in identifying the spectral turnover frequency and providing tighter constraints on the mass loss rate of the progenitor. The only telescope currently planned that could target \snr at 50\,MHz is the low frequency array of the Square Kilometre Array \citep[SKA1-Low;][]{Dewdney2013,2015ExA....39..567D}. However, assuming Equation \ref{eqn:massdensity} provides an accurate evolution of the circumstellar medium and that the turnover in the spectrum is currently $\sim$\,50\,MHz, we would predict a turnover frequency of $\sim$\,10\,MHz by the time SKA1-Low becomes operational in approximately 2023. Hence, it is possible that SKA1-Low will not provide better constraints on the spectral turnover than that presented in this paper. 

Continual monitoring of \snr with the MWA will be useful in investigating the physical properties of the circumstellar medium and physics of diffusive shock acceleration. Gradual flattening of the spectral index, in line with higher frequency observations \citep{2010ApJ...710.1515Z}, will provide a more accurate measure of the strength of the magnetic field in the forward shock and the shock compression ratio \citep{2006ApJ...650L..59B}. Additionally, any observed steepening in the spectral index at low radio frequencies, relative to the high frequency observations, would be indicative of a re-acceleration of electrons by a central compact object \citep{2014ApJ...796...82Z}. A decrease in the spectral index would suggest the interaction of the shock front with a denser circumstellar medium than currently predicted by mass loss models of the progenitor. 

\section{Conclusion}
\label{sec:concl}

We have presented observations of \snr between 72\,MHz and 8.4\,GHz, with the MWA observations representing the lowest frequency observations of \snr to date. This large lever arm in frequency space has allowed us to probe the circumstellar environment of \snr and test different mass loss scenarios of the progenitor at an unprecedented level.

The radio spectrum of \snr does not show any deviation from a non-thermal power-law with a spectral index of $\alpha = -0.74 \pm 0.02$. Since free-free absorption has to cause a spectral turnover to occur below 72\,MHz, we derived an upper limit on the optical depth $\tau_{72} \leq 0.1$, placing an upper limit on the emission measure of $\mathrm{EM} \lesssim 13,000$\,cm$^{-6}$\,pc and electron density of $n_{\mathrm{e}} \lesssim 110$\,cm$^{-3}$. These limits are consistent with limits determined from the detection of the prompt radio emission and X-ray spectra.

The mass loss rate or wind velocity, or both, derived from previous the hydrodynamic simulations were found to be too high to be consistent with our electron density upper limit. The mass loss rate of $\dot{M} \sim 5 \times 10^{-6}$\,$M_{\sun}$\,yr$^{-1}$ and the wind velocity of $v_{\mathrm{w}} = 10$\,km\,s$^{-1}$, derived from optical data, are compatible with our observations. Therefore, we predict a current spectral turnover frequency between $\sim$\,5 and 60\,MHz. We conclude that while SKA1-Low will provide lower frequency observations than reported in this paper, due to the progression of the shock into a lower density circumstellar medium, it will be unlikely SKA1-Low will detect the spectral turnover once it is operational in approximately 2023.

\section*{Acknowledgements}

J.~R.~C. wishes to thank Peter Tuthill for useful discussions about the composition of red supergiant winds, and acknowledges the support of the Australian Postgraduate Award. B.~M.~G. acknowledges the support of the Australian Research Council through grant FL100100114. The Dunlap Institute is funded through an endowment established by the David Dunlap family and the University of Toronto. This scientific work makes use of the Murchison Radio-astronomy Observatory, operated by CSIRO. We acknowledge the Wajarri Yamatji people as the traditional owners of the Observatory site. Support for the operation of the MWA is provided by the Australian Government Department of Industry and Science and Department of Education (National Collaborative Research Infrastructure Strategy: NCRIS), under a contract to Curtin University administered by Astronomy Australia Limited. We acknowledge the iVEC Petabyte Data Store and the Initiative in Innovative Computing and the CUDA Center for Excellence sponsored by NVIDIA at Harvard University. This research was conducted by the Australian Research Council Centre of Excellence for All-sky Astrophysics (CAASTRO), through project number CE110001020. The Australia Telescope Compact Array is part of the Australia Telescope National Facility which is funded by the Commonwealth of Australia for operation as a National Facility managed by CSIRO.




\bibliographystyle{mnras}
\bibliography{SNR_1987A_accepted.bbl}

\bsp	
\label{lastpage}
\end{document}